\documentclass[a4paper,aps,pre,showpacs,twocolumn]{revtex4}
\usepackage[dvips]{graphicx}
\usepackage{amsmath}
\usepackage{amssymb}
\usepackage{psfrag}
\usepackage{color}
\usepackage{natbib}
\usepackage{soul}
\usepackage{wasysym}
\usepackage{pifont}

\newcommand{\red}[1]{{\color{red} #1}}    

\newcommand{\beq}{\begin{equation}}
\newcommand{\eeq}{\end{equation}}

\newcommand{\glv}{\gamma_{lv}}

\newcommand{\rhol}{\rho_\ell}

\newcommand{\upd}{\mathrm{d}}

\newcommand{\Ai}{\mbox{Ai}}
\newcommand{\Aout}{A_{\mathrm{out}}}

\newcommand{\lc}{\ell_c}
\newcommand{\lcurv}{\ell_{\mathrm{curv}}}
\newcommand{\Li}{L_I}
\newcommand{\Lo}{L_O}
\newcommand{\Rfilm}{R_{\mathrm{film}}}
\newcommand{\Rnd}{{\cal R}}
\newcommand{\rtip}{r_{\mathrm{tip}}}
\newcommand{\wo}{\delta}

\newcommand{\woc}{\delta_c}
\newcommand{\tdelta}{\tilde{\delta}}
\newcommand{\sqq}{\sigma_{\theta\theta}}
\newcommand{\srr}{\sigma_{rr}}

\newcommand{\Uindent}{W_{\mathrm{indent}}}
\newcommand{\Ustretch}{U_{\mathrm{stretch}}}
\newcommand{\Ugrav}{U_{\mathrm{gpe}}}
\newcommand{\Usurf}{W_{\mathrm{surf}}}

\renewcommand{\k}{K_f}

\begin{document}

\title{Indentation of ultrathin elastic films and the emergence of asymptotic isometry}  

\author{Dominic Vella$^1$, Jiangshui Huang$^{2,3}$,  Narayanan Menon$^2$, Thomas P. Russell$^3$ and Benny Davidovitch$^2$}
\affiliation{$^1$Mathematical Institute, University of Oxford, Oxford OX2 6GG, United Kingdom\\
$^2$Physics Department, University of Massachusetts, Amherst, Massachusetts 01003, USA\\
$^3$Polymer Science and Engineering Department, University of Massachusetts, Amherst, Massachusetts 01003, USA}

\begin{abstract}
We study the indentation of a thin elastic film floating at the surface of a liquid. We focus on the onset of radial wrinkles at a threshold indentation depth and the evolution of the wrinkle pattern as indentation progresses far beyond this threshold. Comparison between experiments on thin polymer films and theoretical calculations shows that the system very quickly reaches the Far from Threshold (FT) regime, in which wrinkles lead to the relaxation of azimuthal compression. Furthermore, when the indentation depth is sufficiently large that the wrinkles cover most of the film, we recognize a novel mechanical response in which the work of indentation is transmitted almost solely to the liquid, rather than to the floating film. We attribute this unique response to a nontrivial isometry attained by the deformed film, and discuss the scaling laws and the relevance of similar isometries to other systems in which a confined sheet is subjected to weak tensile loads.
\end{abstract}



\maketitle

When an elastic sheet is subjected to external forces, it is often implicitly assumed that the work done is stored in the deformed sheet. Under purely tensile loads,  the  work is stored primarily by stretching energy. When the forces are purely compressive, as in uniaxial buckling, the strain is typically negligible, and the  work is instead stored as bending energy \cite{landau}. Under more complicated compressive forces, such as  those required  to confine a sheet in a  box \cite{witten07}, the  work is stored in localized (stress-focusing) zones that involve bending and stretching. In this Letter, we report a new  response  exhibited by the indentation of an elastic film floating at a liquid--gas interface. We show that for sufficiently large indentations, only a negligible fraction of the work done by the indenter is stored as elastic energy --- the majority is stored in the gravitational and surface energies of the liquid.

\begin{figure}
\centering
\includegraphics[width=0.85\columnwidth]{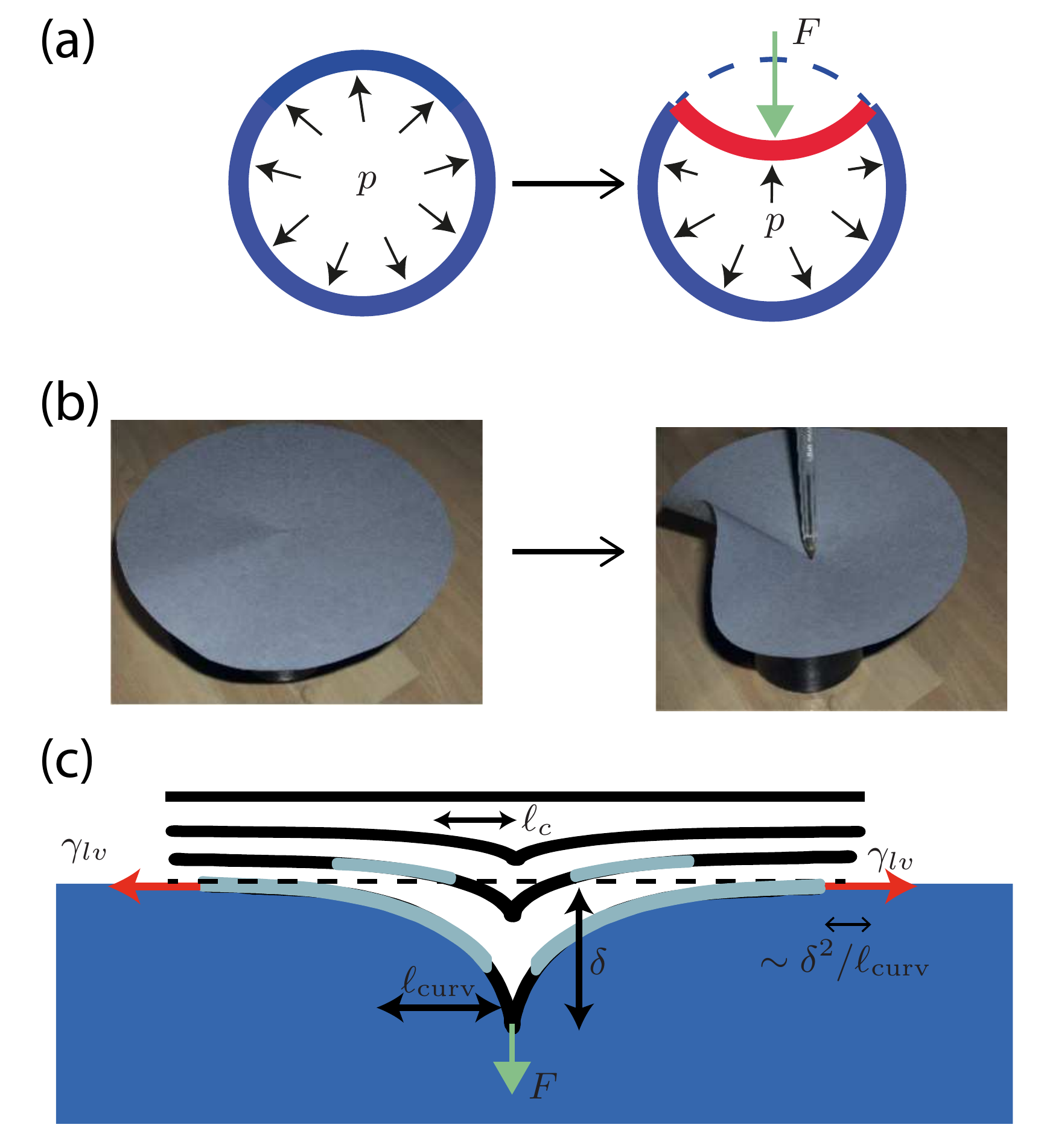}
\caption{(a) Schematic illustration of the  indentation of a very thin axisymmetric  shell with or without an internal pressure, which tends to an isometric  ``mirror buckling" deformation \cite{pogorelov,vella12}.  (b) Indenting a sheet with free boundaries leads to a ``d-cone", which is isometric everywhere except close to the indenter \cite{cerda98,chaieb98}. (c) Schematic illustration showing the evolution of  a floating film subject to increasing  indentation, $\tdelta$: pre-indentation state (upper, flat); at small indentation ($\tdelta \ll \tdelta_c$, second from top) the tension is approximately uniform; at intermediate indentation ($\tdelta_c <\tdelta \ll \Rnd^{2/3}$, second from bottom), the hoop stress is compressive in an annular wrinkled region (light blue); at large indentation ($\tdelta \gg \Rnd^{2/3}$, bottom), wrinkles cover the entire film except for $r<\Li$ (black).}    
\label{fig:setup}
\end{figure}

Interest in the indentation of elastic objects includes a range of metrological applications. Just as one tests an object's stiffness by poking it, controlled indentation is used in the measurement of  internal pressure within  polymeric \cite{gordon04} and biological \cite{HernandoPerez12,arnoldi00,Milani13,arfsten10,vella12} capsules and to determine the modulus of thin membranes \cite{bernal07}. These applications motivated  theoretical studies of indented spherical shells, which suggested that  `mirror-buckling' \cite{pogorelov}  (fig.~\ref{fig:setup}a)  occurs  in the presence of an internal pressure \cite{vella12}. Mirror buckling is the simplest possible isometric (i.e.~strainless) deformation of an infinitely thin shell  so the work done in indenting the shell is nearly independent of the elastic moduli; instead it goes into compressing the gas within the shell \cite{vella12}.

In contrast to shells, the indentation of elastic sheets is highly sensitive to tension. If  a sheet is not under tension, indentation typically leads to the formation of a developable cone (``d-cone") \cite{benamar97,cerda98,chaieb98}, which is isometric everywhere except within a small region around the indenter (fig.~\ref{fig:setup}b). The formation of this nearly isometric shape involves large vertical deflections of the initially planar sheet and is therefore unattainable when vertical displacements are penalized. This is the case for thin elastic films floating on a liquid as formed by vulcanization of a liquid polymer drop, in which case an unknown pre-stress is hypothesized \cite{bernal11}, or by deposition, in which case the liquid surface tension acts at the film's edge \cite{Holmes10}. Experiments on the latter system are better controlled than the former and show that indentation gives rise to a shape full of radial wrinkles that transform into sharp folds  beyond a  threshold indentation \cite{Holmes10}.

The striking difference between the observed wrinkled/folded shape and the nearly isometric d-cone, was interpreted in \cite{Holmes10} as an indication of  considerable strain in the  film induced by the combination of indentation and boundary tension. Here we focus  on the  wrinkle pattern, and show that  wrinkling reveals a new isometry of the film with the strain at the pre-indentation level. As a result, the indentation force exhibits a nontrivial dependence on the surface tension and  density of the liquid, but is independent of the film's elastic moduli. This type of isometry is novel in the elasticity of thin bodies \cite{witten07}, being achieved only in the doubly asymptotic limit of weak applied tension and small bending stiffness; we therefore refer to it as an \emph{asymptotic isometry}.
        
Our experimental setup consists of  polystyrene films (Young's modulus $E=3.4\mathrm{~GPa}$, Poisson ratio $\nu=0.33$ and radius $\Rfilm=1.14\mathrm{~cm}$) floating at the surface of deionized water \cite{huangthesis}. The interfacial  tension, $\glv$, was varied in the range $36\mathrm{~mN/m}\leq\glv\leq72\mathrm{~mN/m}$ using surfactant. The thickness of the film, $t$, satisfied $85\mathrm{~nm}\leq t\leq 246\mathrm{~nm}$ \footnote{See supplementary material at \texttt{http://link.aps.org/supplemental/10.1103/PhysRevLett.114.014301} for further details of experiments and theoretical calculations.}. Stainless steel needles (tip radii  $\rtip\approx 25\mathrm{~\mu m}, 135\mathrm{~\mu m}$) were used to impose a  vertical displacement, $\delta$, at the center of the film. Indentations  up to $\delta\approx 0.75\mathrm{~mm}$ were applied and measured to within $10\mathrm{~\mu m}$. The deformed film was viewed from above using a microscope. 

Our theoretical study is based on the F\"{o}ppl-von Karman (FvK) equations for an elastic film, with stretching modulus $Y = Et$ and bending modulus $B=Et^3/12(1-\nu^2)$, floating on a liquid  of density $\rho_l$, subject to tension $\glv$ at its edge and a localized indentation force $F$ causing a vertical displacement $\wo$ at $r=0$. We assume that the film's radius $\Rfilm$ is much larger than the capillary length $\lc = (\glv/\k)^{1/2}$, where $\k=\rhol g$. 

\begin{figure}
\centering
\includegraphics[width=0.85\columnwidth]{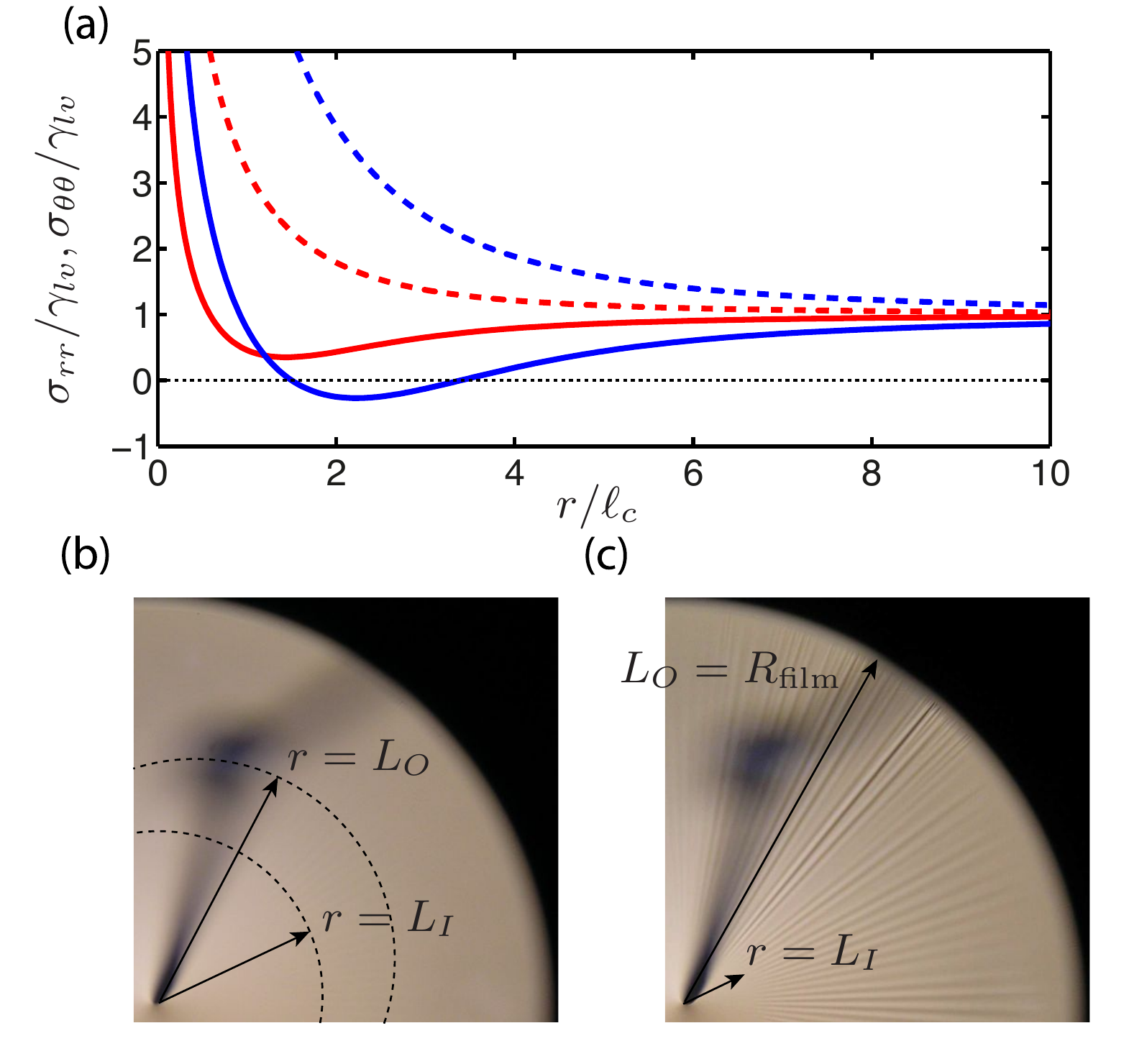}
\caption{(Color online)  The profiles of the hoop (solid curves) and radial (dashed curves) stresses within an indented, unwrinkled film (with $\Rnd\gg1)$ at indentation depths $\tdelta=7.5$ (red curves) and  $\tdelta=15$ (blue curves). Notice that $\sqq $ is negative for intermediate values of $r$ when $\tdelta=15$ so that sufficiently thin films will, in fact, wrinkle.  (b) Just beyond the onset of instability ($\wo=0.48 \mathrm{~mm}$)  wrinkles are confined to an annulus $\Li\leq r\leq \Lo$. (c) Ultimately wrinkles reach the edge of the film (here $\wo=0.56\mathrm{~mm}$) and wrinkles occupy $\Li\leq r\leq \Rfilm$. Here $t=85\mathrm{~nm}$.}
\label{fig:stressplots}
\end{figure}

It is useful to  identify the dimensionless groups in the problem by describing the characteristic behavior of the film as $\delta$ increases (fig.~1c). For very small $\delta$, the response is similar to that of a fluid membrane: the stress remains close to its pre-indentation state, $\srr \approx \sqq \approx \glv$, and the vertical deformation $\zeta(r)$ decays over a horizontal distance $\lc$ (fig.~\ref{fig:setup}c).  As $\delta$ is increased, the indentation-induced strain, $\sim(\delta/\lc)^2$, leads to a noticeable inhomogeneity in the stress (fig.~\ref{fig:stressplots}a): the radial stress $\srr(r)$ decreases monotonically towards $\glv$ for $r \gg \lc$, while the hoop stress $\sqq(r)$ decreases more sharply, overshooting $\glv$ before approaching $\glv$ from below. Intuitively, this occurs because indentation causes material circles to be pulled inwards and become relatively compressed. If the indenter's tip is sufficiently small, this purely geometric effect is  governed only by the ``confinement ratio" $(\delta/\lc)^2 / (\glv/Y)$ between the indentation-induced strain, and the purely tensile strain caused by surface tension. We therefore introduce the dimensionless indentation depth
\beq
\tdelta=\frac{\wo}{\lc}\left(Y/\glv\right)^{1/2} \ , 
\label{eqn:deltatilde}
\eeq 
which determines the stress profiles fully. As $\tdelta$ increases above a threshold $\tdelta_c$, analysis of the FvK equations shows that the hoop stress becomes compressive ($\sqq<0$) within a narrow annulus (blue solid curve, fig.~2a). As these films are very thin, a compressive hoop stress causes wrinkling (fig.~2b). For a film with infinite radius, numerical analysis of the FvK equations yields $\tdelta_c \approx 11.75$, in  good agreement with our  experiments for a range of film thicknesses, tensions and indenter sizes (fig.~3a). 

Two crucial phenomena occur as the indentation amplitude is increased above  $\tdelta_c$. First, the wrinkled zone expands: a detailed calculation \cite{InPrep} shows that the outer radius of the wrinkled zone  $\Lo/\lc\sim \tdelta^{3/2}$ so that wrinkles reach the film's edge when $\tdelta \sim (\Rfilm/\lc)^{2/3}$. Second, the thinness of the film means that the compressive hoop stress is \emph{completely relaxed} by wrinkling: $\sqq(r)\approx0$, a qualitative change from the prebuckled and compressive (but unwrinkled) profiles [the solid red and blue curves, respectively, in Fig.~2a). We therefore use the Far-from-Threshold (FT) approach, valid in the singular limit of zero bending stiffness \cite{stein61,pipkin86}. These two phenomena are characterized by the dimensionless radius, $\Rnd$, and ``bendability", $\epsilon^{-1}$, \cite{davidovitch11} of the film, where: 
\beq
\Rnd=\Rfilm/\lc,\quad\epsilon^{-1}=\glv\lc^2/B.
\eeq For our experiments, $\epsilon\lesssim 10^{-5}$.

In the FT approach  the energy is written $U = U_{\rm dom} + U_{\rm sub}$ 
with $U_{\rm sub}$ the subdominant energy governed by the bending cost of wrinkling, which vanishes as $\epsilon \to 0$, and $U_{\rm dom}$ the dominant energy, which remains finite as $\epsilon \to 0$. Minimization of $U_{\rm sub}$ determines the  number of wrinkles. In the current study we employ tension field theory \cite{stein61} (minimizing $U_{\rm dom}$) to determine the mean deflection profile $\zeta(r)$ and the extent of the wrinkles.

We write the axisymmetric FvK equations using an Airy potential $\psi$ (so that $\srr=\psi/r$ and $\sqq=\psi'$). The vertical force balance reads
\beq
 B\nabla^4\zeta-\frac{1}{r} \frac{\upd}{\upd r}\left(\psi \frac{\upd\zeta}{\upd r}\right) =-\k \zeta - \frac{F}{2\pi r} \delta(r), 
 \label{eqn:fvk1}
\eeq
where $F$ is the point-like indentation force, found as part of the solution for a given indentation. The compatibility of strains in the unwrinkled zone (where both $\srr$ and $\sqq$ are tensile) gives \cite{landau}
\beq
r\frac{\upd }{\upd r}\left[\frac{1}{r}\frac{\upd }{\upd r}\left(r\psi\right)\right]=-\frac{1}{2}Y\left(\frac{\upd \zeta}{\upd r}\right)^2 \ .
\label{eqn:fvk2} 
\eeq We note that  equations \eqref{eqn:fvk1}-\eqref{eqn:fvk2} are invariant under $\zeta\to-\zeta$, $F\to-F$; our results therefore apply equally to the cases of pushing down on (considered here) and pulling up on \cite{Holmes10} a floating membrane. Invoking tension field theory, we neglect the bending term in Eq.~(\ref{eqn:fvk1}), and replace Eq.~(\ref{eqn:fvk2})  by $\psi = {\rm constant}$ in the wrinkled zone (since $\sqq = 0$) \cite{endnote35}. 

\begin{figure}
\centering
\includegraphics[width=0.8\columnwidth]{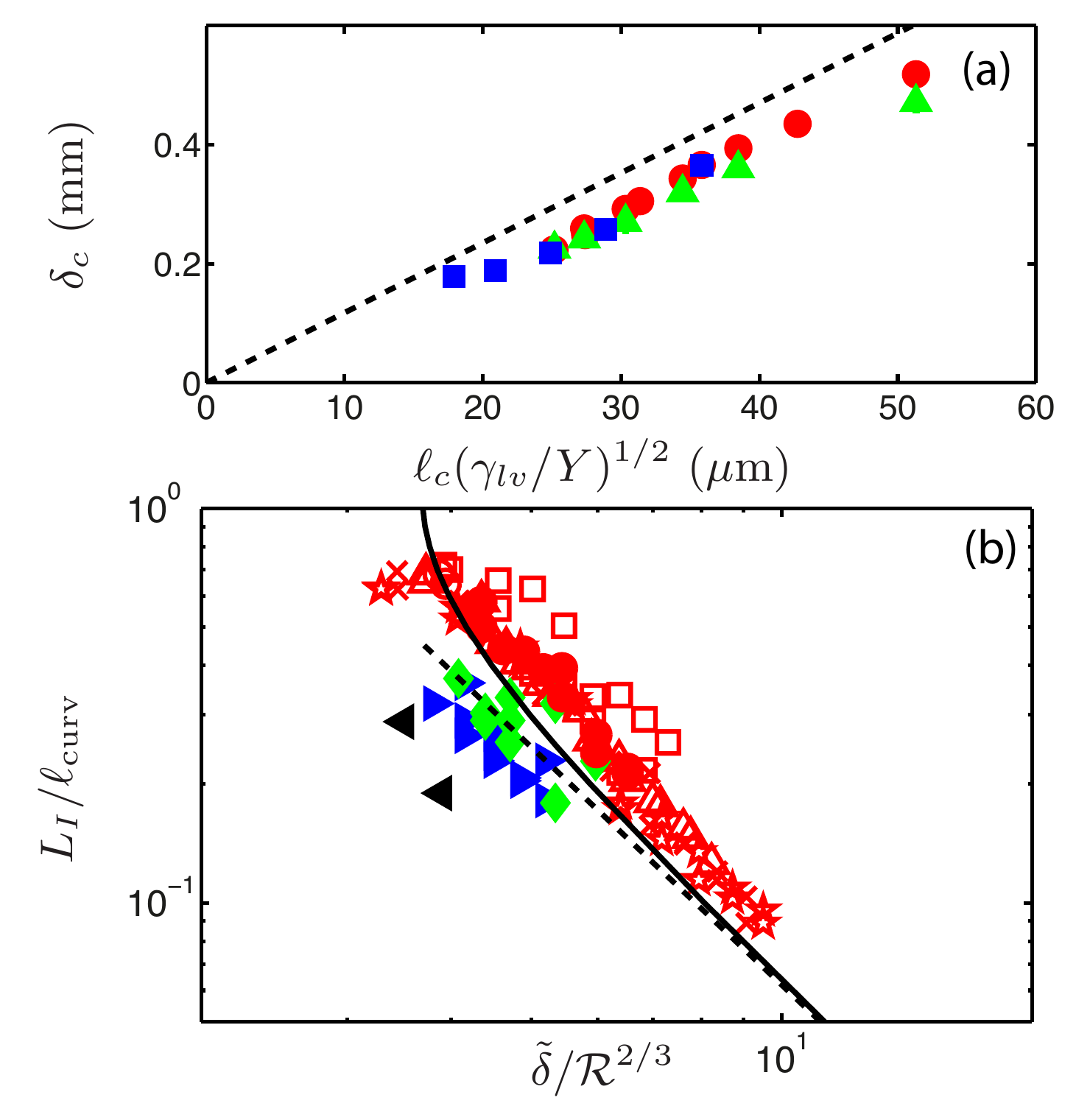}
\caption{(Color online) (a) Experimentally measured threshold indentation for wrinkling, $\woc$, as a function of $\lc(\glv/Y)^{1/2}$. Experiments  with varying film thickness  ($59\mathrm{~nm}\leq t\leq 246\mathrm{~nm}$), $\glv=72\mathrm{~mN/m}$ are shown for indenter radii $\rtip=135\mathrm{~\mu m}$  ({\color{green} $\blacktriangle$}) and $\rtip=25\mathrm{~\mu m}$ ({\color{red} {\large $\bullet$}}). Experiments with varying surface tension coefficient  ($36\mathrm{~mN/m}\leq\glv\leq72\mathrm{~mN/m}$) and $t=121\mathrm{~nm}$ ({\color{blue} $\blacksquare$}). The theoretical prediction for $\Rnd\gg1$, $\tdelta_c\approx11.75$, is also shown (dashed line). Good agreement with experiment justifies our neglect of  indenter size and any (hypothesized) manufacture-dependent pre-stress, which were both attributed  crucial roles previously \cite{bernal11}.
(b) The  inner wrinkle radius, $r=\Li$, decreases with increasing indentation, $\tdelta$, when  wrinkles reach the film's edge. Experiments with $\glv=72\mathrm{~mN/m}$ and: $t=85\mathrm{~nm}$ ($\red{\Box}$), $t=121\mathrm{~nm}$ ($\red{\bigcirc}$), $t=158\mathrm{~nm}$ ($\red{\bigtriangleup}$), $t=207\mathrm{~nm}$ ($\red{\times}$), $t=246\mathrm{~nm}$ (\red{\ding{73}}). Experiments with  $t=121\mathrm{~nm}$ and: $\glv=58\mathrm{~mN/m}$ (${\color{green} \blacklozenge}$), $\glv=50\mathrm{~mN/m}$ (${\color{blue} \blacktriangleright}$) and $\glv=42\mathrm{~mN/m}$ ($ \blacktriangleleft$). The  prediction of the FT theory (solid curve) and the asymptotic result  \eqref{eqn:smalllamasy} (dashed line) are also shown. The wrinkle number scales similarly to that found in other studies \cite{huang07} (data not shown).}
\label{fig:wrinkprops}
\end{figure}

We turn now to large indentations $\tdelta \gg \Rnd^{2/3}$, where the wrinkles cover the whole film except in  $0<r<L_I$ (see fig.~\ref{fig:stressplots}c). Noting that $\srr(\Rfilm) = \glv$, and that $\sqq \to 0$ in the wrinkled zone, we find that $\srr (r) = \glv \Rfilm/r$ for $L_I<r<\Rfilm$;  Eq.~\eqref{eqn:fvk1} then reduces to Airy's equation  \cite{abramowitz64}: 
\beq
\zeta(r)  = \Aout \cdot \Ai (r/\lcurv), \quad
\lcurv =\Rnd^{1/3} \lc.
\label{eq:lcurvFT3}
\eeq 
Here $\lcurv$, which increases with film size $\sim \Rfilm^{1/3}$, replaces $\lc$ as the decay length of membrane deflections.

The prefactor $\Aout$ and the inner radius, $\Li$, are found by patching the wrinkled zone to the unwrinkled core ($r<L_I$). In the limit $\tdelta \gg \Rnd^{2/3}$ we find, using standard techniques \cite{bhatia68,chopin08,vella10,endnote35}, that $\Aout \approx -\delta/\Ai(0)$ and the radial displacement at the edge of the film approaches a limiting value:      
\beq
u_r(\Rfilm) \approx -0.243~\wo^2/\lcurv \ ;
\label{eq:urRfilmFT3}
\eeq a result whose importance will become apparent shortly. Our asymptotic calculations also reveal that
\beq
\frac{\Li}{\lcurv}\approx6.20 ~\left(\tdelta/\Rnd^{2/3}\right)^{-2} 
 \  \Rightarrow  \ \Li \sim \frac{\Rfilm^{5/3}\glv^{5/3}}{Y\k^{2/3}}\ \wo^{-2} \ . 
\label{eqn:smalllamasy}
\eeq
At the scaling level, Eq.~(\ref{eqn:smalllamasy}) can be understood by noting that in the tensile core the indentation-induced radial stress $\sim Y (\wo/\lcurv)^2$, whereas in the wrinkled zone $\srr = \glv \Rfilm/r$. Continuity of the radial stress at $r=\Li$ yields the scaling in \eqref{eqn:smalllamasy}. Figure~\ref{fig:wrinkprops}b shows that this result agrees  well with numerical solutions of the full problem  and with experiments. Strikingly, Eq.~(\ref{eqn:smalllamasy}) shows that the size of the tensile core is affected by all physical parameters in the problem (except the bending modulus).         

Our calculation also yields the indentation force $F\approx 4.581 \glv\Rnd^{2/3}\wo$, consistent with previous measurements  \cite{endnote35,Holmes10}. Two features of the scaling $ F \sim \glv^{2/3}K_f^{1/3} R^{2/3} \wo $, are surprising. Firstly, $F\propto\wo$, even though the FvK equations are highly non-linear. Secondly, the force is independent of the elastic moduli of the film. Understanding this mechanical response requires reconsideration of the dominant energy of the wrinkle pattern: 
\beq
U_{\rm dom} = - (\Uindent + \Usurf) +  (\Ugrav  + \Ustretch).
\label{eq:FT-dom-energy}
\eeq   
Here $\Uindent, \Usurf$ are the work done by the indentation force and surface tension acting at the edge of the film, respectively.  $\Ugrav,\Ustretch$ are the gravitational energy of the displaced liquid and the elastic energy of the film, respectively. The work done by the indentation force $\Uindent=\int F~\upd\delta\sim \glv \Rnd^{2/3}\wo^2$. One might assume that $\Uindent$ would be transmitted to the elastic energy $\Ustretch$ due to the tensile components of the compression-free stress field. However, integrating the strain energy density $\sigma_{ij}^2/Y$ we obtain:    
$\Ustretch/\Uindent \sim (\tdelta/\Rnd^{2/3})^{-2} \ll 1$. Indeed, using Eqs.~(\ref{eq:lcurvFT3},\ref{eq:urRfilmFT3}) to evaluate the work $\Usurf \sim \Rfilm \glv u_r(\Rfilm)$ of the surface tension, and the energetic cost $\Ugrav\sim \k\int_0^{\Rfilm}  \zeta^2r~\upd r$ of the vertically-displaced liquid, we find the asymptotic relation: 
\beq 
{{\rm for} \ \epsilon^{-1} \gg 1 ,   \tdelta \gg \Rnd^{2/3} :}  \  \  \Uindent  \to  - \Usurf +  \Ugrav  
\label{eq:asym-iso}
\eeq         

This energetic structure describes a novel mechanical response of an elastic film, whereby the work of the indenter is transmitted predominantly to the subphase (increasing gravitational energy and uncovering surface area of the liquid), while an asymptotically negligible fraction is stored  as elastic energy in the film. This simple energetic structure reflects a nontrivial geometric feature: the wrinkled film becomes isometric to its pre-indentation state in the {\emph{doubly asymptotic}} limit of small bending modulus ($\epsilon \ll 1$) and small exerted tensile strain (since $\tdelta \gg \Rnd^{2/3} \Rightarrow (\glv/Y) \ll u_r(\Rfilm)/\Rfilm$ by eqn \eqref{eq:urRfilmFT3}). In this doubly asymptotic limit, the hoop strain $\epsilon_{\theta\theta}$ is and asymptotic isometry follows from Eqs.~(\ref{eq:lcurvFT3},\ref{eq:urRfilmFT3}), which yield the elimination of radial stretching in the limit $\tdelta \Rnd^{-2/3} \to \infty$ (the apparent stretching, $\sim \sqrt{\wo^2 + \lcurv^2} - \lcurv$, is completely cancelled by the lateral displacement $u_r(\Rfilm)$ of the edge). Thus, the formation of wrinkles at negligible energetic cost enables the metric of the film to remain almost identical to its pre-indentation state, even though the film suffers a large deflection, Eq.~(\ref{eq:lcurvFT3}), that is determined by  indentation, gravity, and surface tension. In other words, the  film lies in a ``no-man's-land" -- too stiff to be stretched significantly (since the applied tensile strain $\glv/Y$ is  small), and yet perfectly deformable (since the bending modulus $B$ is also small).

In conclusion, we have shown that an indented floating film starts with a purely tensile response but evolves, with the aid of wrinkles, into a state that is asymptotically isometric to its initial state. This demonstrates a novel mechanical response, in which the indenter does work mainly on the liquid with only a negligible fraction transmitted to the elastic film. This response also underlies the stability of the poked film to two common failure modes of floating objects: the film would sink if the displacement at the edge exceeds $\lc$ \cite{vella15}, but $\zeta(R_{\rm film})  \propto \delta \Ai(\Rnd^{2/3}) \ll \lc $ (since $ \Rnd \gg 1$). Similarly, pulling-induced delamination will occur if the adhesive energy, $\Delta\gamma\Rfilm^2$, is smaller than the alternative deformation energy \cite{wagner11}. Here, the alternative deformation energy,  $U_{\rm dom}$, is barely affected by the elastic modulii of the sheet, so delamination is expected only  for $\delta > \sqrt{\Delta\gamma/\gamma_{lv}}\Rfilm^{2/3}\lc^{1/3}$, which is beyond the reach of existing experiments \cite{Holmes10} and the validity of our small slope theory.

The concept of asymptotic isometry should be relevant to other systems, where a thin elastic object is forced into a curved, nondevelopable shape, in the presence of weak tensile loads. Representative examples include the wetting of a film by a liquid meniscus \cite{Py07,King12} or its adhesion to a sphere \cite{Hure11,hohlfeld14}, and the twisting of a stretched ribbon \cite{Chopin13,Chopin14}. Such systems may also  have parameter regimes in which the object is  highly deformed yet nearly isometric to its undeformed state; consequently, the work done by external forces is not stored in the object itself. Finally, it is important to realize that  asymptotically isometric states may not necessarily be wrinkled: the wrinkle-fold transition \cite{Holmes10,pineirua13} and other secondary instabilities may also exhibit a similar phenomenology. We hope that our work will provide a suitable framework for studying these phenomena.                  

\acknowledgments

We thank E.~Hohlfeld for insightful discussions on asymptotic isometry. This research was supported a Leverhulme Trust Research Fellowship (D.V.), the W.M.~Keck Foundation, and NSF Materials Research Science and Engineering Center at UMass Amherst DMR-0820506 (J.H., T.P.R., N.M., B.D.), NSF-CAREER award DMR 11-51780 (B.D.) and NSF DMR 1207778 (N.M.).

\end{document}